\documentclass[12pt]{article}
\usepackage{bbm}
\usepackage{amssymb}
\newcommand{\nc}[2]{\newcommand{#1}{#2}}
\nc{\Z}{\mathbbm{Z}}
\nc{\N}{\mathbbm{N}}
\nc{\R}{\mathbbm{R}}
\nc{\T}{\mathbbm{T}}
\nc{\C}{\mathbbm{C}}
\nc{\D}{\mathbbm{D}}
\nc{\A}{\mathbbm{A}}
\nc{\mP}{\mathbbm{P}}
\nc{\cA}{{\cal A}}
\nc{\cF}{{\cal F}}
\nc{\cO}{{\cal O}}
\nc{\cH}{{\cal H}}
\nc{\cP}{{\cal P}}
\nc{\cG}{{\cal G}}
\nc{\fN}{\frak{N}}
\nc{\id}{{\bf 1}}
\nc{\beq}{\begin{equation}}
\nc{\eeq}{\end{equation}}
\nc{\bea}{\begin{eqnarray}}
\nc{\eea}{\end{eqnarray}}
\nc{\beas}{\begin{eqnarray*}}
\nc{\eeas}{\end{eqnarray*}}
\nc{\ba}{\begin{array}}
\nc{\ea}{\end{array}}
\nc{\lra}{\longrightarrow}
\nc{\ra}{\rightarrow}
\nc{\ot}{\otimes}
\nc{\ci}{\circ}
\nc{\bpf}{{\em Proof:\enspace\enspace}}
\newtheorem{pr}{Proposition}
\newtheorem{de}{Definition}
\newtheorem{them}{Theorem}
\newtheorem{lem}{Lemma}

\newtheorem{re}{Remark}

\nc{\bpr}{\begin{pr}}
\nc{\bth}{\begin{them}}
\nc{\ble}{\begin{lem}}
\nc{\bco}{\begin{corollary}}
\nc{\bre}{\begin{re}}
\nc{\bex}{\begin{example}}
\nc{\bde}{\begin{de}}
\nc{\ede}{\end{de}}
\nc{\esa}{\end{satz}}
\nc{\epr}{\end{pr}}
\nc{\ethe}{\end{them}}
\nc{\ele}{\end{lem}}
\nc{\eco}{\end{corollary}}
\nc{\ere}{\hfill\mbox{$\Diamond$}\end{re}}
\nc{\eex}{\end{example}}
\nc{\epf}{\hfill$\square$}
\nc{\Dom}{{\rm Dom}}
\nc{\Ran}{{\rm Ran}}
\nc{\tpsi}{\tilde{\psi}}
\nc{\tphi}{\tilde{\varphi}}
\nc{\teta}{\tilde{\eta}}

\begin{document}

\title{NONLINEAR DIRAC EQUATIONS  and
NONLINEAR GAUGE TRANSFORMATIONS}

\author{H. - D. DOEBNER\\Department of Physics, Institute for Theoretical
Physics,\\ Technical
University of Clausthal, Germany,\\
E-Mail: asi@pt.tu-clausthal.de\\[.3cm]
R. ZHDANOV\\
Institute of  Mathematics,\\
Ukrainian National Academy of Sciences,\\ Kiev
}
\date{}
\maketitle
\begin{abstract}
\noindent
Nonlinear Dirac equations (NLDE) are derived through a group $\fN^2$
of nonlinear (gauge) transformation acting in the corresponding state
space. The
construction generalises a construction for nonlinear Schr\"odinger
equations. To 
relate $\fN^2$ with physically motivated principles we assume: locality 
(i.e. it
contains
no explicit derivative and no derivatives of the wave function),
separability (i.e. 
it acts on product states componentwise) and Poincar\'e invariance.
Furthermore 
we want that a positional density is invariant under $\fN^2$. Such nonlinear 
transformations yield NLDE which describe physically equivalent systems. To
get 'new' 
systems, we extend this NLDE (gauge extension) and present a family of NLDE 
which is a slight nonlinear generalisation of the Dirac equation. We
discuss and 
comment the fact that nonlinear evolutions are not consistent with the
usual 
framework of quantum theory. To develop a corresponding extended framework
one needs models for nonlinear evolutions which also
indicate possible physical
consequences of nonlinearities.
\end{abstract}

\newpage

\section{INTRODUCTION\label{in}}

There is some recent interest in fundamental nonlinear quantum mechanical 
evolution equations. Different proposals for (non relativistic) nonlinear 
Schr\"odinger equations (NLSE) are known: Some of them are physically plausible 
modifications of the linear equation (LSE) \cite{bm76}--\cite{k-twb78}; 
others are based on some first 
principles (non relativistic quantum mechanics) \cite{dg92}--\cite{dgn96}. 
It is known \cite{dg96}--\cite{p-j} (see
\cite{dg03} 
for a short review) that nonlinear evolutions are not consistent with the usual 
(linear) quantum mechanical framework. For an interpretation and application of 
NLSE a corresponding extension of this framework is needed \cite{p-j,cd02}.
If this 
extension allows to show in approximation, e.g., with a nonlinear evolution
equation, 
that a nonlinearity yields measurable effects, one can decide whether a deviation from the linear prediction can be viewed as an 
information on a nonlinear extension of quantum mechanics
\cite{gkz81}--\cite{dms00}.\\[.3cm]
The recent results refer to non relativistic Schr\"odinger equations. A 
generalisation to relativistic Dirac equations is possible. One can use the 
method \cite{dg96,dgn96} of nonlinear transformations in the state space;
they are assumed to be consistent with first 
principles: Locality and separability condition, Poincar\'e invariance and a 
notion for an equivalent description of quantum systems. We present such a 
generalisation with a step-by-step construction (see a detailed review
\cite{dz03}). A 
family of nonlinear Dirac equations (NLDE) is obtained which is 'near' or 
'similar' equivalent to those families which are equivalent to linear
ones.\\[.3cm]  
We explain our construction in section 2. Families of nonlinear
transformations are constructed 
in section 3 with the resulting NLDE and their extensions.
Section 4 contains concluding remarks.

\section{A METHOD TO CONSTRUCT NONLINEAR DIRAC EQUATIONS   (NLDE)}

      Consider the function space $\cG_4=C^1(\C^4,\C)$ 
             on Minkowski space-time, i.e. 
complex (column) vector valued functions $\psi$   with components                     
$\psi_j=\psi_j(x_0,\ldots,x_3),~j=0,1,2,3;$  define
$\bar{\psi}=\psi^*\gamma_0$. The Hilbert space of a 1-particle 
Dirac system is denoted as $\cH^1_D$; use $\tilde{\cG}_4=\cG_4\cap\cH^1_D$
 for an interpretation of the 
later results. The evolution is given through a family $\cF_0$
 of  linear Dirac 
operators and equations  (parameter $m\geq 0$) 
\beq
\D_D=(\gamma_\mu p^\mu-m),~\D_D\psi=0\mbox{   with
}p^\mu=i\frac{\partial}{\partial x^\mu} \mbox{   on  } \tilde{G}_4.
\eeq      

Consider a group $\frak{N}$  of invertible (in general nonlinear) 
transformations $N$   
  acting on $\cG_4$  or $\tilde{\cG}_4$
\[
N:\psi\mapsto N\psi.
\]
The $N$   may depend on $\psi$   (and $\bar{\psi}$), derivatives of $\psi$,
$\bar{\psi}$ and explicitly on  $\partial_\mu$  
and $x_\mu$. If $\psi$ is a solution of  $\D_D\psi=0$ then  $\psi'=N^{-1}\psi$
is a solution of
\[
  \D_D^N\psi'=0\mbox{          with     } \D^N_D=\D_D N.
\]           

$\D_D^N$    is a (linearisable) nonlinear Dirac operator.

Our construction of a physical acceptable nonlinear Dirac operator starts with 
the group $\frak{N}$. As mentioned (section \ref{in}) we select 
subgroups of $\frak{N}$  which are consistent with plausible additional requirements 
('first principles') in step 1 - 4. In step 5 we extend the obtained families of 
NLDE to families   not equivalent to  $\cF_0$. 

\begin{itemize}
\item[]
STEP 1 (Locality):

\item[]
The Dirac operator $\D_S$ is a first order PDO. We want the same for
$\D_S^N$. 
This  implies a locality condition: $N$ does not depend on derivatives 
of $\psi$, $\bar{\psi}$    and not explicitly on  $\partial_\mu$. 
For later invariance properties (step 4) 
it is reasonable to assume that also an explicit dependence of $x_\mu$ (and 
functions of $x_\mu$) will not appear. Hence one has
\beq\label{np}                                      
N\psi=N(\psi)\psi=(N_0(\psi)\psi_0,\ldots,N_3(\psi)\psi_3)^\top.
\eeq                                                                              
 
It is convenient to use also   $N\psi\equiv
N[\psi]\equiv
N(\psi)\psi$. The set of transformations (\ref{np})  is a 
group $\frak{N}^l\subset\frak{N}$;  the $N(\psi)$ act on $\cG_4$ 
as matrix valued multiplication operators.

\item[]
STEP 2 (Separability)
\item[]
Quantum theory describes not only 1-particle systems but necessarily also 
systems build from $n$ particles. The corresponding $n$-particle Hilbert space      
 is - following first principles -  a product space of 1-particle spaces; it is 
spanned through the linear completition of a set of product wave functions
\[
\cP^{(n)}=\{\psi^1\ot\cdots\ot\psi^n~|~\psi^i\in\cH^1_D, i=1,\ldots,n\}.
\]

To extend linear operators $\A\equiv\A^{(1)}$ on $\cH^1_D$ to (linear) 
operators  $\A^{(n)}$ on $\cH^n_D$ one defines first an action 
on $\cP^{(n)}$,
\[
\A^{(n)}(\psi^1\ot\cdots\ot\psi^n)=\A\psi^1\ot\cdots\A\psi^n.
\]

This definition extends uniquely by linear completion from $\cP^{(n)}$
to $\cH^n_D$ (or a 
dense set in $\cH^n_D$). For the nonlinear operators - like those considered here -  
this construction for $n$-particle operators from 
1-particle ones is not possible. To have a property of  $N^{(n)}$
which is at least 
partly consistent with the linear theory we assume that $N$ acting on 
$\cH^1_D$ extend 
to the set $\cP^{(n)}$    as
\beq\label{npr}
N^{(n)}(\psi^1\ot\cdots\ot\psi^n)=N\psi^1\ot\cdots\ot N\psi^n,
\eeq                                                                            
                                                           
which is a (weak) separation property for $N$. The set of $N$
with (\ref{npr}) forms a  
subgroup   $\frak{N}^s$    of  $\frak{N}$. Note that one needs additional 
information (see e.g. \cite{p-j}) to extend   $N^{(n)}$ from $\cP^{(n)}$ to $\cH^n_D$. 
This above construction yields evolution equations for 1-particle systems only.

\item[]
      STEP 3  (Poincar\'e Invariance)
\item[]
The Dirac Operator $\D_D$ in $\cH^1_D$ behaves under the inhomogeneous 
Lorentz group  with 
a spin $\frac{1}{2}$ representation $U$  in $\cH^1_D$. This invariance is 
a principal property of 
the Dirac system. However, for the transformed operator $\D_D^N$ this 
property may be 
lost. Therefore one has to guarantee it through an assumption on $N$. 
For this we 
use the following information: The generators of $U$ are
\[
\mP_\mu=p_\mu;~~J_{\mu\nu}=x_\mu p_\nu-x_\nu p_\mu+\frac{i}{2}(\gamma_\mu
\gamma_\nu
-\gamma_\nu \gamma_\mu).
\]

$N$ is a function of $\psi$ (step 1). 
This leads easily to a condition      

\beq\label{nu}
N(U\psi)=UN(\psi)   \mbox{   with    }                
U=\exp\sum_{\mu,\nu=0,\mu\neq\nu}^3\alpha_{\mu\nu}\gamma_\mu \gamma_\nu,
\eeq           

which implies Poincar\'e invariance of $\D_D^N$. The set of                     
local   tranformations in$\frak{N}^l$ with condition (\ref{nu})  
form a group $\frak{N}^{l,P}$.
\item[]
      STEP 4 (Equivalence)
\item[]
In non relativistic quantum mechanics (e.g., in $\R_x^3$)
one can argue that the  positional density 
$\rho(x,t)$ for all $x$ and $t$ is a fundamental observable of a system  
in the sense 
that all observables can be calculated from the information encoded  in      
$\rho(x,t)$. This density is connected to (pure) states $\varphi$ through 
$\varphi^*\varphi$. 
and was called utility function in \cite{dz03} . For transformations  $N$ with an invariant 
utility function 
\beq\label{nph}
N[\varphi]^*N[\varphi]=\varphi^*\varphi,
\eeq     

the  positional densities (\ref{nph}) for $\varphi$  and for  $N^{-1}\varphi$ 
are equal. This implies a notion 
of equivalence. (\ref{nph}) is an equivalence condition. 
Wave functions $\psi$ which evolve with  a  
Schr\"odinger operator
\[
\D_S=i\partial_t-H
\]
              
($H$ a Hamiltonian) and a transformed wave function $N^{-1}\varphi$ with Schr\"odinger operator 
(which may be nonlinear) $\D_SN$ describe the same physics; they are
equivalent.
Linear $N[\varphi]$, i.e. $N[\varphi]=\exp ia(\vec{x},t)\varphi$ are usual
gauge transformations. It is reasonable to denote  general
$N$ in the following as nonlinear gauge transformations \cite{dg96,dgn96}.\\[.3cm]
In the relativistic Dirac case a corresponding argument is used. Here the 
utility  function  with a physically reasonable density $\rho$ is      
\beq\label{fs}
F_\rho={\psi}^\dagger\psi.
\eeq                                                        
with $\psi^\dagger=(\psi_0^*,\ldots,\psi^*_3)$.                                            
Dirac systems related through  $N$  are equivalent if $N$ leaves 
$F_\rho$ invariant, i.e.
\beq\label{inv}
N[\psi]^\dagger N[\psi]={\psi}^\dagger\psi.
\eeq
Transformations $N\in\frak{N}^l$ with (\ref{inv}) form a group
$\frak{N}^{l,e}$.

\item[]
      STEP 5 (Extension)
\item[]
The  intersection
\beq                                        
\fN^{l,s}\cap\fN^{l,P}\cap\fN^{l,e}=\fN^2
\eeq

yields a family  $\cF_1$ of NLDE which is - because of the invariance of the 
utility function - equivalent to the linear family $\cF_0$. With steps 1.-4.  one 
obtains interesting nonlinear reformulations of the linear Dirac equation.  
However, we are looking for 'new' systems which are 'near' to 'physically
equivalent' ones, i.e. in systems which are local, separable and Poincar\'e
invariant 
but not equivalent. The following construction leads to such systems 
through a simple 'extension' of $\cF_1$:  It turns out that $\fN^2$ is 
characterised by a real number and through complex functions which are related 
among themselves but otherwise arbitrary. The evolution equations depend on the 
relation between these functions. Hence one can 'extend' the family $\fN^2$
if one 
breaks this relation, i.e. if one chooses the functions and the 
number independently.  Such a family describes 'new' systems. We remark already 
here that a general framework to treat spin $1/2$ particles with nonlinear 
evolutions is not (yet)  known and a NLDE is not immediately applicable.
\end{itemize}

\section{RESULTS OF THE CONSTRUCTION}

We explain details of the above step-by-step method and present some obtained 
NLDE families.

\subsection{LOCALITY AND SEPARABILITY\label{lose}}

In step 1 we motivated the subgroup $\fN^l$ of nonlinear transformations 
which are 
functions  of $\psi$ and which act as $4\times 4$ matrices on
$\cG_4$.

To fulfil  the separation property (\ref{npr}) in step 2 it is sufficient to 
discuss 
the two particle case n=2,
\beq\label{sep}
N^{(2)}(\psi^1\ot\psi^2)=N\psi^1\ot N\psi^2,~~\psi^i\in \cH^1_D,~i=1,2,
\eeq
i.e. we demand the existence of an $n$ component function $N^{(2)}$   
on  
such that for $N$ and for any two $\psi^1$    and $\psi^2$
     relation (\ref{sep}) holds. For $n>2$ we get corresponding results.
To
calculate the resulting form of  $ N$ use the (non unique) polar 
decomposition
 of the components   $\psi_i^k=R_i^k\exp iS_i^k,\;\; k=1,2,\; i=0,\ldots
,3$,
express the components 
of $N(\psi^k)\psi^k$             in terms of $R^k, S^k$   
with $R^k=\{ R_1^k, \ldots, R_n^k \}$, $S^k=\{ S_1^k, \ldots, S_n^k \}$
and write 
\[
N(\psi^k)\psi^k=F(R^k,S^k), ~~k=1,2.
\]
To ensure (\ref{sep}) we have to prove the existence of functions $G_{ij}$
such that
\begin{equation}
\label{sep1}
F_i(R^1, S^1)F_j(R^2, S^2)=G_{ij}({\cal R},
{\cal S}),  
\end{equation}
where $R^1,  S^1,  R^2, S^2$ have $4$ components, ${\cal R}, {\cal S}$ 
have the
components $R^1_iR^2_j$ and $S^1_i + S^2_j$ and $i,j=0,\ldots,3$.
The $G_{ij}$
are invariant
under the following one-parameter groups with real parameters 
$\tau,$ $\theta$
\begin{eqnarray*}
&1.& \mbox{The group of scale transformations on ${\cal{R}}$}\\ 
&&{ R}^{1'}={
R}^1\exp(\tau ),\quad {R}^{2'}={ R}^2\exp(-\tau),\\
&2.&\mbox{The group of translations on ${\cal{S}}$ }\\
&&{S}^{1'}={S}^1+\theta ,\quad {S}^{2'}={S}^2-\theta 
\end{eqnarray*}
which are 
generated by  Lie vector fields
\[
D=\sum_{j=0}^3\left(R_j^1{\partial\over\partial R_j^1}-
R_j^2{\partial\over\partial R_j^2}\right),\quad
\mbox{\rm and}\quad 
P=\sum_{j=0}^3\left({\partial\over\partial S_j^1}-
{\partial\over\partial S_j^2}\right).
\]
Take the functional equation (\ref{sep}) and
apply $D$ and $P$; 
the right hand side vanishes and 
yields the following system of
differential-functional equations:
\[
(DF_i)F_j + (DF_j)F_i=0,\quad (PF_i)F_j + (PF_j)F_i=0,
\]
$i,j=0,\ldots,3$. Dividing these equations by
(non-zero) functions $F_iF_j$ we represent them as follows:
\begin{eqnarray}
&&{1\over{F_i}}\sum_{k=0}^3R_k^1{\partial F_i\over\partial R_k^1}=
{1\over{F_j}}   
\sum_{k=0}^3R_k^2{\partial F_j\over\partial R_k^2},\\
&&{1\over{F_i}}\sum_{k=0}^3{\partial F_i\over\partial S_k^1}=
{1\over{F_j}}   
\sum_{k=0}^3{\partial F_j\over\partial S_k^2}
\end{eqnarray}  
for  $i,j=0,\ldots,3$.
Both
sides of the above equations depend on
different variables, hence there exist complex parameters
$a$ and $b$ such that 
\begin{eqnarray}
&&\sum_{k=0}^3R_k\frac{\partial F_i}{\partial R_k}=aF_i,\label{sep2}\\
&&\sum_{k=0}^3\frac{\partial F_i}{\partial S_k}=bF_i \label{sep3}
\end{eqnarray}
holds for any $i=0,\ldots,3$.

For $n>1$ the general solution of (\ref{sep2}) can be written in the form
\[
F_i=(R_i)^{a}   
H_i\left(\frac{R_1}{R_0},\ldots,\frac{R_{3}}{R_0},
S_0,\ldots,S_3\right),\quad i=0,\ldots,3.
\]
with arbitrary smooth complex-valued functions $H_i$; instead of
$\frac{R_j}{R_0},~j=1,2,3,$ one can also use $\frac{R_j}{R_k},~j\neq k,~
k\mbox{ fixed,} ~k,j=0,\ldots,3$. 
With (\ref{sep2})
\[
\sum_{k=0}^3\frac{\partial H_i}{\partial S_k}=b,
\quad i=0,\ldots,3
\]
holds.
Integrating the equations  we arrive at the final form\footnote {Instead of $R_k/R_3, ~k=0,\ldots, 2$ one can use also
$R_k/R_l, ~l  \mbox{ fixed}, ~ k=0,\ldots, 3,~k\neq l$, e.g., $l=0$.} for
$F_i$
\begin{equation}
F_i=(R_i)^a\exp(b S_i)
G_i\left(\frac{R_1}{R_0},\ldots,\frac{R_{3}}{R_0},
S_1-S_0,\ldots,S_{3}-S_0\right).
\label{Fi}
\end{equation}  
Here $G_i$ are arbitrary smooth functions of the indicated 
variables and $i=0,\ldots,3$.
We
suppose 
that the functions $H_i$ are well-defined under
$R_3\to 0$.
This form of $N$
is  necessary  for the separation condition of $N^{(n)}$ on 
${\cal{P}}^{(n)}$ for any $n\geq 2$.
It is also sufficient. For a straightforward proof use 
the identities for $n=2$
\begin{eqnarray*}
&&\frac{R^1_i}{R^1_n}=\frac{R^1_iR^2_j}{R^1_nR^2_j},\quad
\frac{R^2_i}{R^2_n}=\frac{R^2_iR^1_j}{R^2_nR^1_j},\\
&&S_i^1-S_n^1=(S_i^1+S_j^2)-(S_n^1+S_j^2),\\
&&S_i^2-S_n^2=(S_i^2+S_j^1)-(S_n^2+S_j^1),
\end{eqnarray*}
Analogous identities are used for  $n>2$.
We summarize the result:\\ 
The general form of a transformation $N$:
\[
N:\cG_4\rightarrow \cG_4
\]
which satisfies locality and separation conditions is given through
\begin{eqnarray}
&&N_{j[a,b,{G}]}(\psi)
=(R_j)^{a-1}\exp((b-i) S_j)
\label{sep4}\\
&&\times G_j\left(\frac{R_1}{R_0},\ldots,\frac{R_{3}}{R_0},
S_1-S_0,\ldots,S_{3}-S_0\right),\nonumber
\end{eqnarray}  
$N$ is labelled by two arbitrary complex parameter $a, b$ 
and $4$ functions $(G_0, \ldots , G_3)\equiv G$.
  The  $G_j$   can be written also in terms of   $\psi, {\bar{\psi}}$    
\begin{equation}
G_j=G_j\left(\frac{\psi_1}{\psi_0},\ldots,\frac{\psi_{3}}{\psi_0},
\frac{{\bar{\psi}}_1}{{\bar{\psi}}_0},\ldots,\frac{{\bar{\psi}}_{3}}
{{\bar{\psi}}_0}
\right)
\label{G}
\end{equation}
We expect that  the
transformations (\ref{sep4}) form a group.  
Arrange the labelling complex parameters
$a$ and $b$ in  matrix form
\begin{equation}
K=K(a,b)=\left(\matrix{\tilde{a}, & \hat{a}\cr
\tilde{b}, & \hat{b}\cr}\right), ~~ a=\tilde{a}+i\hat{a},
~~b=\tilde{b}+i\hat{b}
\label{matrix}  
\end{equation}
and compute the product 
\[
N_{[K_1,{ G}_1]}\ci
 N_{[K_2,{ G}_2]}=N_{[K_3,{ G}_3]},
\]
or, more detailed,
\begin{equation}
N_{[K_2,{G}_2]}(N_{[K_1,{G}_1]}(\psi)\psi )N_{[K_1,{G}_1]}
(\psi)\psi=
N_{[K_3,{G}_3]}(\psi)\psi.
\label{NKG}
\end{equation}
The result for $K_3$   is the matrix product
\[
K_3=K_2K_1.
\]
For the components of  $G_3$ we find
\begin{equation}
G_{3j}=|G_{1j}|^{a_2}\exp(b_2argG_{1j})G_{2j}(u_1,\ldots,u_{3},
v_1,\ldots,v_{3})
\label{G3}
\end{equation}
the variables of $G_{1j}$     were given in (\ref{sep4}); the $u_l$,
$v_l$      in $G_{2j}$     are ($l=1,2,3$)
\[
u_l=\left(\frac{R_l}{R^0}\right)^{\tilde{a}_1}\exp \tilde{b}_1
(S_l-S_0)|G_{1l}||G_{10}|^{-1},
\]
\[
v_l=\hat{a}_1\ln\left(\frac{R_l}{R^0}\right)+ \hat{b}_1
(S_l-S_0)+\arg(G_{1l}-G_{10})
\]
Hence the transformations (\ref{sep4}) build a local (infinite parameter) 
group $\fN^{l,s}$
 for  $|K|\neq 0$
             and for appropriate  ${G}$. The element  $N_{[{\bf 1},
1]}$
              is
the identity;   $N_{[K,G]}$     
     is locally invertible in a neighbourhood of the
identity; the associativity is respected. 


\subsection{POINCAR\'E INVARIANCE}     

We gave in (\ref{nu}) a condition for $N\subset \fN^l$ which yields a 
Poincar\'e invariant 
nonlinear Dirac operator. A straightforward evaluation of (\ref{nu}) 
 (see \cite{nz}, Theorem 1.2.1; \cite{nz1}), yields the most general form 
for $N[\psi]$,
\beq\label{gen}
N(\psi)\psi=(f_1(\bar{\psi}\psi,\bar{\psi}\gamma_5\psi)+
f_2(\bar{\psi}\psi,\bar{\psi}\gamma_5\psi)\gamma_5)\psi.
\eeq                                                                                         
$f_1$, $f_2$ are independent complex functions depending on the invariant 
quantities         
$\bar{\psi}\psi\equiv X,~~\bar{\psi}\gamma_5\psi\equiv Y$ ($X$ depends on $R_k^2$,
$Y$ on $R_iR_k$ and $S_i-S_k$.). The result reflects that  $N$    
transforms under the Poincar\'e group like the corresponding scalar and 
pseudoscalar invariants. The invariant transformations $N\in\fN^l$ form a 
group  $\fN^{l,P}$.

We are interested now in local, separable and Poincar\'e invariant $N$, i.e., 
in
\[
\fN^{l,s,P}=\fN^{l,s}\cap\fN^{l,P}.
\]
The intersection is  given ( see (\ref{sep4}) and (\ref{gen})) through
\beq\label{int}
N_{j[a,b,G]}(\psi)\psi_j=R_j^{a-1}\exp(b-i)S_jG_j(.,.)\psi_j=
f_1(X,Y)\psi_j+f_2(X,Y)(\gamma_5\psi)_j
\eeq
e.g. for $j=0$     as
\beq\label{ra0}
R^a_0\exp(b-i)S_0G_0(.,.)=f_1(X,Y)R_0+f_2(X,Y)R_2\exp i(S_2-S_0).
\eeq
This relates $G_j(.,.)$  and  $f_1,~ f_2$, which depend 
on $R_i, ~S_i$. Similarly as in section  \ref{lose}
we use groups of translations in ${\cal S}$
 and  of scale transformation on ${\cal R}$,
\[
S_j\mapsto S_j+\theta,~~~R_j\mapsto \lambda R_j.
\]
The variables in $G_j(.,.)$ are invariant under both 
groups; the variables in $f_1$, $f_2$, i.e. $X$ and $Y$, are translation 
invariant
and behave under scale transformation as
$X\mapsto\lambda^2X,~Y\mapsto\lambda^2Y$. 
The translations imply for  $j=0$ (and any $j$) 
in (\ref{ra0})
\beq\label{bbb}
b=i,\mbox{ i.e., } \tilde{b}=0,~~\hat{b}=1.
\eeq
For scale transformations we use in $f_1,f_2$ instead of $X$  and $Y$ the 
variables $X$      
and $Z=\frac{X}{Y}$ ($XY\neq 0$) and get 
\[
R_0^aG_0(.,.)=\lambda^{-a+1}F(\lambda^2X,\ldots),~~~
F(\lambda^2 X,\ldots)=f_1(\lambda^2X,Z)R_0+f_2(\lambda^2X,Z)R_2\exp
i(S_2-S_0).
\]
Differentiate this in respect to $\lambda$ and obtain a differential equation;
 its 
solution  specify for $\lambda=1$ the $X$   and $Y$   dependence of $f_1$ 
and  $f_2$;
\beq\label{dep}
f_i(X,Y)=g(X)h_i(Z), ~~g(X)=(\bar{\psi}\psi)^\frac{a-1}{2},
~~h_i(Z)\mbox{ arbitrary },i=1,2.
\eeq
or with
$X=\bar{\psi}\psi=R_0^2+R_1^2-R_2^2-R_3^2$
\beq\label{gx}
g(X)=(R_0^2+R_1^2-R_2^2-R_3^2)^\frac{a-1}{2}.
\eeq
For the relation between  $G_j(\ldots)$ and  $h_1,h_2$ we get 

\beq\label{dot}
G_j(\ldots)=\left(\frac{R_0^2+R_1^2-R_2^2-R_3^2}{R_j^2}\right)^\frac{a-1}{2}
(h_1(Z)-ih_2(Z)\frac{R_{J(j)}}{R_j}\exp i(S_{J(j)}-S_j))
\eeq
($J$ maps $0\mapsto 2$, $1\mapsto 3$, $2\mapsto 0$, $3\mapsto 1$.)
$Z$ depends on $R_j,S_j$ as $R_jR_k, S_j-S_k$.

Nonlinear transformations  (\ref{int}) with (\ref{dep}) 
leads to NLDE which are local, separable and 
 Poincar\'e invariant.  They build a group $\fN^1\equiv\fN^{l,s,P}$; 
their elements are labelled 
through $a,~h_1(Z),~h_2(Z)$
and act on $\cG_4$ as
\beq\label{nah}
N_{[a,h_1,h_2]}(\psi)\equiv
(\bar{\psi}\psi)^\frac{a-1}{2}\cdot(h_1(Z)+h_2(Z)\gamma_5).
\eeq

\subsection{INVARIANCE OF THE UTILITY FUNCTION}

The utility function  $F_\rho$ was defined in (\ref{fs}) - step 4 - and 
those  $N$   with
invariant $F_\rho$
%
define equivalent systems.  This equivalence condition is an additional 
restriction to the
different subgroups of $\fN$.  We discuss first the case $\fN^{l,s}$
       and specialise the 
result to  $\fN^1=\fN^{l,s,P}$.\\[.3cm]  
CASE   $\fN^{l,s}$:\\  
Insert $N_{j[a,b,G]}(\psi)\psi$ from ( \ref{sep4}   )  in the equivalence
condition (\ref{inv})
\begin{equation}
\sum_{k=0}^3R_k^{2\tilde{a}} \exp\tilde{b} S_k|G_k|^2=
\sum_{k=0}^3R_k^2
\label{R}
\end{equation}
As in the last section use  translations in ${\cal S}$ and 
scale transformations in ${\cal R}$. Because  (\ref{R}) 
is invariant under both types  one gets  
\beq\label{bt}
\exp 2\tilde{b}\theta\cdot\sum_{k=0}^3R_k^{2\tilde{a}}\exp(2\tilde{b}S_k)
|G_k|^2=
\sum_{k=0}^3R_k^{2}
\eeq
\[
\lambda^{2\tilde{a}-2}\cdot\sum_{k=0}^3R_k^{2\tilde{a}}\exp(2\tilde{b}S_k)
|G_k|^2=\sum_{k=0}^3R_k^2.
\]
From this we have immediately  
\[
\tilde{a}=1,~ \tilde{b}=0;~ 
\hat{a}   \mbox{ and }\hat{b}     
\mbox{ arbitrary},
\] 
and the  $G_k$ are restricted to those $G_k'$
             which
fulfil
\begin{equation}
\sum_{k=0}^3R_k^{2}(|G_k'|^2-1)=0
\label{R2}
\end{equation}
i.e. the $G_k'$ with $k=0,1,2,3$ are not independent.

\noindent
Hence, a transformation $N$
which satisfies locality, separation and equivalence conditions is given
through
\[
N_{j[K',{ G}']}\psi_j=R_j^{1+i\hat{a}}\exp i\hat{b}S_jG_j'
\left(\frac{R_1}{R_0},\ldots,\frac{R_{3}}{R_0},
S_1-S_0,\ldots,S_{3}-S_0\right)
\]
with
\begin{equation}
K'=\left(\matrix{1, & \hat{a}\cr
0, & \hat{b}\cr}\right), \;\;\hat{a}, \hat{b}\in\R \; \mbox{arbitrary,}\;
\hat{b}\neq 0,~~~ 
\sum_{k=0}^3R_k^{2}(|G_k'|^2-1)=0.
\label{matrix1}
\end{equation}

The restriction in (\ref{matrix1}) from  $K$  to $K'$    and from $G_k$    
to $G_k'$      is
enforced through the equivalence condition.

The transformations $N_{[k',G']}$ form the group $\fN^{l,s,e}$.
For $N_{[k_3',G_3']}=N_{[k_1',G_1']}\ci N_{[k_2',G_2']}$ we find for $K_3'$
\begin{equation}
K_3'=K'_2K'_1=\left(\matrix{1, & \hat{a}_1+\hat{a}_2\hat{b}_1\cr
0, & \hat{b}_1\hat{b}_2\cr}\right)
\label{matrix2}
\end{equation}
and for the $G'_{3j}$, $j=0,1,2,3$     the following condition (from (\ref{G3}), 
(\ref{gx}), (\ref{dot}): 
\begin{equation}
|G_{3j}'|^2=|G_{1j}'|^2|G_{2j}'|^2, 
~~~\sum_{k=0}^3R_k^{2}(|G_{1k}'|^2|G_{2k}'|^2-1)=0 
\label{GG}
\end{equation}
with $G_j'$ and its variables from (\ref{sep4}), (\ref{G3}).
An explicit form of $G_{3,j}$ is given in the next example.
\newpage
\noindent
Case $\fN^1=\fN^{l,s,P}$:\\ 
Local, separable and Poincar\'e invariant $N_{[a,h_1,h_2]}$ 
in (\ref{nah})
are special cases 
of $N_{[a,b,G]}$ in (\ref{sep4}).
           Therefore the equivalence condition (\ref{inv}) leads to 
(\ref{R}) and we
have together with the invariance condition (\ref{bbb})
\beq\label{aa}
a=1+i\hat{a},~~(b=0).
\eeq
If we insert $G_j'(.,.)$ from (\ref{dot}) in (\ref{R}) we have with
(\ref{matrix2}) 
\[
\sum_{k=0}^3R_k^2(|h_1(Z)-ih_2(Z)\frac{R_{J(k)}}{R_k}\exp
i(S_{J(k)}-S_k)|^2-1)=0.
\]
This is a condition for   $h_i(Z)$, $i=1,2$, which are functions of
$Z=\frac{X}{Y}$ and implies
\[
|h_1(Z)|^2+|h_2(Z)|^2=1.
\]

Hence we have
\beq\label{cosi}
h_1(Z)=\exp i\phi(Z)\cos\rho(Z),~~~h_2(Z)=\exp i\phi(Z)\sin\rho(Z),
\eeq
with two real functions  $\phi(Z),\rho(Z)$ which label $N_{[a\phi,\rho]}$
together with a real
number  $\hat{a}$;     they     act on $\cG_4$    as
\bea\label{naf}
N_{[\hat{a},\phi,\rho]}(\psi)\psi&=&(\bar{\psi}\psi)^{i\frac{\hat{a}}{2}}
\exp i(\cos\rho(Z)+i\sin\rho(Z)\gamma_5)\psi\\
&=&\exp(i\frac{\hat{a}}{2}\ln(\bar{\psi}\psi)+i\phi(Z)+\rho(Z)\gamma_5)\psi.
\nonumber
\eea
Again, the  $N_{[\hat{a},\phi,\rho]}$ form a group $\fN^2=\fN^{l,s,P,e}$. 
The group relation is
\[
N_{[\hat{a}_1,\phi_1,\rho_1]}\ci N_{[\hat{a}_2,\phi_2,\rho_2]}=
N_{[\hat{a}_3,\phi_3,\rho_3]}
\]
with
\[
\hat{a}_3=\hat{a}_1+\hat{a}_2,~~\rho_3(z)=\rho_1(z)+\rho_2(\tilde{z}),
~~\tilde{z}(z)=\frac{z\cos 2\rho_1(z)-\sin 2\rho_1(z)}
{\cos 2\rho_1(z)+z\sin 2\rho_1(z)},
\]
\[
\phi_3(z)=\frac{\hat{a}_2}{2}\ln\left(\cos 2\rho_1(z)-\frac{1}{z}\sin
2\rho_1(z)\right)+\phi_1(z)+\phi_2(\tilde{z}).
\]
$|G_{j,3}|^2$ is in this example independent of $j$ and
$|G_{j,3}|^2=1$
holds; condition (\ref{R2}) is fulfilled.

\subsection{THE RESULTING NLDE AND THEIR EXTENSIONS}

Part of our construction is that invertible $N$ imply for $N^{-1}\psi$
a nonlinear
Dirac operator  and equation
\beq\label{dir}
(\gamma_\mu p^\mu-m)N({\psi})\psi=(\gamma_\mu p^\mu+H(\psi))\psi=0
\eeq
with $H(\psi)$ as nonlinear term depending on $m$ and the labels of $N$.
%
%
%
We calculate the  NLDE for $N\in\fN^1,\fN^2$.\\
%
In case of $N\in\fN^1$ we find for $H_1$
($\frac{dh_i(Z)}{dZ}=h_i'(Z)~,~i=1,2$) from (\ref{nah})
\bea
H_1(X,Y)&=&(\gamma_\mu p^\mu
X)\frac{a-1}{2X}\nonumber\\
&&+(\gamma_\mu p^\mu Z)
\frac{h_1(Z)h_1'(Z)+h_2(Z)h_2'(Z)+
\left(h_1(Z)h_2'(Z)-h_1'(Z)h_2(Z)\right)\gamma_5}{h_1(Z)^2+h_2(Z)^2}
\nonumber\\
&&+m\frac{h_1(Z)^2-h_2(Z)^2+2h_1(Z)h_2(Z)\gamma_5}{h_1(Z)^2+h_2(Z)^2}.
\label{h1}
\eea
This is a family $\cF_1$ of Poincar\'e invariant NLDE which
respect the separability condition.\\
For the interesting subfamily
$\cF_2\subset \cF_1$ which describes in addition physically equivalent 
systems we find
with $h_i(X)$ in (\ref{cosi}) and with (\ref{aa})
\bea
H_2(X,Y)&=&(\gamma_\mu p^\mu
X)i\frac{\hat{a}}{2X}\nonumber\\
&&+
(\gamma_\mu p^\mu Z)\cdot\left(i\phi'(Z)+
\rho'(Z)\gamma_5)\right)-m\exp2\rho(Z)\gamma_5.\label{h2}
\eea
$\cF_2$ is invariant under $\fN^2$, i.e. if one calculates  
$\D_D^{N_1}\ci N_2$, $N_{1,2}\in\fN^2$ one gets a 
nonlinear term of the form (\ref{h2}).  
\\
The structure in (\ref{h1}), (\ref{h2}) stems from the 
general form of
local, Poincar\'e invariant $N$ in (\ref{gen}). The operator
$\gamma_\mu p^\mu$ acts on the
invariants $X$, $Z$ and the coefficient functions have a scalar and 
pseudoscalar part. Solutions  $\psi'$  for $\cF_1$ 
are available from solutions $\psi$
  of $\D_S\psi=0$ through
$\psi'=N^{-1}\psi$, $N\in \fN^1$ or $\fN^2$. By construction,
the family  $\cF_2$ describes equivalent systems . \\
To get `new'
systems we use the method explained in Section 2, step 5. We generalize
$H_2(X,Y)$ with real functions $g(X)$, $k(Z)$, $l(Z)$, $n(Z)$ to

\bea
H_{2,ex}(X,Y)&=&(\gamma_\mu p^\mu
X)ig(X)\nonumber\\
&&+
(\gamma_\mu p^\mu Z)\cdot\left(ik(Z)+
l(Z)\gamma_5)\right)-m\exp2n(Z)\gamma_5.\label{h2e}
\eea
The resulting `extended' family $\cF_2^{ex}$
is
local and Poincar\'e invariant; furthermore it is
not
equivalent to $\cF_0$   but it has a `similar' structure. $\cF_2$
is a
subfamily of $\cF_2^{ex}$ with
\beq
k(Z)=\phi'(Z),~~l(Z)=\rho'(Z),~~n(Z)=2\rho(Z).
\eeq
 To obtain solutions of (\ref{h2e}) the above mentioned technique 
could
be useful.
Because $\cF_2$ is constructed from $\fN^2$, i.e. from a family of nonlinear
gauge functions (cf. Section 2, step 5), $\cF^{ex}_2$ is denoted as gauge
extension of $\cF_0$.

\section{CONCLUDING REMARKS}    
We derived from physically reasonable `principles' a family $\cF^{ex}_2$
of NLDE for
one-particle systems which is a `mild' (gauge) extension of the family
$\cF_2$ which
describes the linear Dirac family $\cF_0$  through a nonlinear 
Dirac operator.
With `mild' we understand that the deviation of $\cF^{ex}_2$ from $\cF_2$
 or the effect of a nonlinearity  $H_{2,ex}$  compared to $H_2$ is `very small';
 this assumes that 
we work
in a region in which very small nonlinear corrections behave well and that
some information is known on a solution variety of $\cF_2^{ex}$.   
\\[.3cm]
Concerning the physical relevance of a family of NLDE we mention the
following arguments:
\begin{itemize}
\item[1.]Because of the successes of the (linear) formulation and
interpretation of quantum theory and the fact that a linear framework does
not allow nonlinear evolution of one particle systems like $\cF^{ex}_2$
  one has to
develop a new framework together with an interpretation if one wants to
incorporate such evolutions. Furthermore one has to show that
nonlinearities yields experimental effects which are measurable through
precision experiments. We have no corresponding results for a nonlinear
Dirac theory (and a connected quantum field theory). However we showed,
using parts of the linear framework and interpretations - denoted as
`first principles' - that one can construct physically equivalent NLDE 
which yield
after gauge-extensions a family of physically motivated one particle NLDE.
                          
For the generalisation  $\cF^{ex}_2$ to $n$-particle systems one has to 
extend the
evolution operator, which is defined only on product states, to a dense set
in $\cH_D^2$ (operator-extension). In the nonrelativistic case such a method is
known \cite{p-j}.
\item[2.] 
A reasonable attempt to develop a framework for a nonlinear quantum
theory should start with the dynamic of the system, i.e. with a physical
justification of a nonlinearity, e.g. from `first principles' or from
geometrical properties \cite{dg92}, \cite{dg94}, \cite{dg96}, \cite{dgn96}.
Such a framework depends on the structure of the
nonlinearity. Therefore some information on a method to derive a NLDE with
a special class of nonlinear terms like $H_2$ could be useful.\\   
\item[3.]
We mentioned in section 1 some recent interest in nonrelativistic
nonlinear quantum mechanical evolutions equations, e.g. nonlinear
one-particle Schr\"odinger equation (NLSE). It would be interesting to see
whether NLSE appear as nonrelativistic limits from a NLDE, as it is the
case for linear Schr\"odinger and Dirac equations. One can realise the
nonrelativistic case with the program from section 1; take a Hilbert
space of scalar functions, the (nonrelativistic) utility function (\ref{nph}) 
and
use the central extension of the inhomogeneous Galilei group (with time
translations) instead of the Poincar\'e  group as space-time invariance. The
result is known (see \cite{dg94}, section 4); the nonlinear terms depend on 
second
order terms $\partial^2_k\varphi,~\partial_k\varphi\partial_l\varphi,~
k,l=1,2,3$ because the Schr\"odinger equation is of second order. 
The usual procedure to derive a nonrelativistic
limit of the Dirac equation leads to the Pauli equation (or the Schr\"odinger
equation for vanishing electromagnetic potentials). If one tries an
analogous procedure for the NLDE family  $\cF_2^{ex}$,   second order terms 
in $\psi, \bar{\psi}$ are absent.
Hence for a NLDE with a nonrelativistic limit given through a NLSE of the
type given in \cite{dg92} - \cite{dgn96} one should use a nonlinear transformation $N$ which
depends also on derivatives of $\psi$. Such transformations 
were discussed in
the nonrelativistic case in \cite{gs02}. 
\end{itemize}
{\bf Acknowledgements}\\[.3cm]
We are very grateful to Dr. Olena Roman for a compilation of some results
in a first version of this paper and to Prof. A.G. Nikitin, Dr. Rainer Matthes 
and Alois Kopp for discussions and help.

\end{document}